\documentclass{aa}
\usepackage{natbib}
\usepackage{graphicx}
\usepackage{amssymb}
\usepackage{color}
\usepackage{amsmath}
\newcommand{\bl}[1]{\mbox{\boldmath$ #1 $}}

\begin{document}

\title{Improving the thin-disk models of circumstellar disk evolution. The 2+1-dimensional model}

\author{Eduard I. Vorobyov \inst{1,2,3} and  Yaroslav N. Pavlyuchenkov \inst{4}}
\authorrunning{Vorobyov et al.}

\institute{Institute of Fluid Mechanics and Heat Transfer, TU Wien, 1060, Vienna, Austria; 
\email{eduard.vorobiev@univie.ac.at}
\and
Research Institute of Physics, Southern Federal University, Stachki Ave. 194, Rostov-on-Don, 
344090 Russia;
\and
University of Vienna, Department of Astrophysics,  Vienna, 1180, Austria
\and 
Institute of Astronomy of the Russian Academy of Sciences, Pyatnitskaya str. 48, 
Moscow, 119017 Russia
}

\abstract
  {Circumstellar disks of gas and dust are naturally formed from contracting pre-stellar molecular 
  cores during the star formation process. 
  To study various dynamical and chemical processes that take place in 
  circumstellar disks prior to their dissipation and transition to debris disks, 
  the appropriate numerical models capable of studying the long-term disk chemodynamical 
  evolution are required.}
  {We aim at improving the frequently used two-dimensional hydrodynamical model for disk evolution 
  in the thin-disk limit by employing  a better calculation of the disk thermal balance and 
  adding a reconstruction of the disk vertical structure.
  Together with the hydrodynamical processes, the
  thermal evolution is of great importance since it influences
  the strength of gravitational instability and the chemical evolution of the disk.}
  {We present a new 2+1-dimensional numerical hydrodynamics model of 
  circumstellar disk evolution, in which the thin-disk model is complemented with 
   the procedure for calculating the vertical distributions of gas volume density and
  temperature in the disk. The reconstruction of the disk vertical structure is performed at every time
  step via the solution of the time-dependent radiative transfer equations coupled to the
  equation of the vertical hydrostatic equilibrium.}
  {We perform a detailed comparison between circumstellar disks produced with our
  previous 2D model and with the improved 2+1D approach. The
  structure and evolution of resulting disks, including the differences in 
  temperatures, densities, disk masses and protostellar accretion rates, are discussed in detail. }
  {The new 2+1D model yields systematically colder disks, while the in-falling
  parental clouds are warmer. Both effects
  act to increase the strength of disk gravitational instability and,
  as a result, the number of gravitationally bound fragments that form in the disk via gravitational
  fragmentation as compared  to the purely 2D thin-disk simulations with a 
  simplified thermal balance calculation.
  The presented method has a low time overhead as compared to the purely 2D models and is 
  particularly suited for the long-term simulations of circumstellar disks including compact chemical reaction networks.}
\authorrunning{Vorobyov \& Pavlyuchenkov}
\titlerunning{The 2+1D model for disk evolution}
\maketitle

\section{Introduction}
Circumstellar disks hold the key to our understanding of stellar mass accumulation and planet formation.
They form thanks to the conservation of angular momentum of 
a collapsing cloud core when the in-spiralling material 
hits the centrifugal barrier near the stellar surface before landing onto the star. 
As the core collapse continues, 
new layers of infalling material are added to the newborn disk at progressively 
larger radial distances, causing the disk to grow in both mass and size. Numerical 
simulations and observations indicate that this process starts in the Class 0 phase 
of stellar evolution and continue until the the parental core depletes or dissipates \citep{Machida2010,
Tobin2015}.
In this so-called protostellar or embedded phase, disks are usually most massive owing to
continuing mass loading from the parental core and are often
prone to the development of gravitational instability  \citep{VB2005,Tsukamoto2013,Lomax2014,Dong2016,Mayer2016}.
This is also the phase when dust growth and perhaps the first phases of planet assembly
are likely to take place \citep{ALMA2015}.

In the subsequent T Tauri phase of stellar evolution, the disk slowly loses its mass owing 
to accretion onto the host star and expands in reaction to angular momentum redistribution 
within the disk.  In this phase, dust growth from small grains to planetesimals and planetary cores
takes place, finally leading to the emergence of protoplanets covered with primordial 
atmospheres \citep[e.g.,][]{Benz2014} and  basic chemical ingredients are converted into
complex (organic) species \citep{Semenov2013}. Finally, the combined action of stellar accretion, 
planet formation, disk winds and photoevaporation leads to the dispersal of the disk 
gaseous component, revealing a debris disk consisting of solids which are left 
from the planet formation process.

Many aforementioned phenomena during the entire disk evolution are controlled by 
the radiative input from the host star and external environment; 
both largely determine the temperature in the disk 
upper layers and set the minimum temperature in its midplane. 
In the steady-state disk models, the radiative input can be taken into account rather
accurately using sophisticated ray-tracing or Monte-Carlo techniques 
\citep[e.g.][]{Woitke2009, RADMC,Andes2013}.
However, due to high computational costs, these models lack a dynamical aspect, 
which can be important when considering the young disks prone to gravitational instability 
or more evolved disks wherein 
planetesimals/planetary cores are subject to growth and migration.
To properly consider these effects, fully dynamical models are needed which are usually based
the equations of hydrodynamics complemented with a module that takes
the radiative input from the stars and/or external environment into account.

In the full 3D hydrodynamics simulations, the radiative input can be taken into account by solving the
equations of radiation transfer, usually with simplifying assumptions such as frequency 
integrated opacities and diffusion approximation \citep[e.g.,][]{Klahr2005,Tsukamoto2015}, 
because the more accurate frequency-dependent models \citep[e.g.,][]{Kuiper2010} are 
computationally expensive.   Even with these simplifying assumptions,
3D models are an inconvenient tool to make a parameter space study of the disk evolution 
for many model realizations and many orbital periods. For this purpose, 2D thin-disk
models of disk evolution have been routinely employed \citep[e.g.][]{Vor2010, Zhu2012, Regaly2013, 
Markus2014} thanks to their low computational costs.
In these models, the equation for thermal balance is usually complemented with 
cooling and heating functions which take the radiative input and local disk cooling into 
account. The form of these cooling and heating terms may differ slightly from study to study,
but they are essentially hinged on the calculated midplane and surface temperatures of the disk and
the optical depth from the disk surface to the midplane. The midplane temperature is usually set 
equal to the hydrodynamic temperature, the surface temperature is calculated assuming the 
black-body character of the incident stellar and background radiation, and the optical depth is 
calculated assuming the vertically isothermal temperature distribution.

While having advantages in simplicity and low computational costs, 
this approach has obvious weaknesses. First, it is not clear if the 2D hydrodynamic
temperature is indeed representative of the disk midplane temperature. Second,
this method  provides little information on the disk vertical structure, essentially assuming 
that the disk is vertically isothermal, which may not be the case. For instance, passively 
irradiated disks are known to exhibit positive temperature gradients in the vertical direction
\citep[e.g.,][]{Dullemond:2002},
while dense gaseous clumps and spiral arms in gravitationally unstable disks 
may be characterized by a more complicated vertical temperature distribution \citep{Vorobyov:2014}.
To circumvent this difficulty, various forms of the vertical density distribution 
can be adopted (Gaussian, exponential), but this procedure cannot be easily applied to the 
vertical temperature distribution. As a result, the 2D models have limited applicability 
to studying, e.g., chemical reactions which are known to sensitively depend on the 
gas temperature (which is likely to be varying in the vertical direction as well).

This paper  presents a method that addresses the aforementioned weaknesses by means of  
coupling the gas dynamics computations in the thin-disk limit and calculations of the 
disk vertical structure. 
The gas dynamics in the disk plane is calculated using the hydrodynamics equations,
while the disk vertical structure is calculated using the  equations of radiation transfer 
and hydrostatic balance  in the vertical direction. As a result, we retrieve the full three-dimensional
density and temperature structure of the disk at every time step, 
which is lacking in purely two-dimensional models. 
The presented method  has low time overhead as compared to the purely 2D thin-disk models
{and it is faster than fully 3D models since the disk gravitational
potential is found using a convolution theorem (see Equation~\ref{gravpot}),
which is not applicable to the fully 3D models formulated
in the curvilinear coordinate systems. The
adopted 1D radiative transfer in the vertical direction is also 
much faster than the fully 3D method.}
Our method is therefore well suited for the long-term simulations of circumstellar disks.
We compare the disk evolution calculated using this 2+1D approach with purely 2D simulations
and briefly discuss the applicability of our new method to calculating the chemical evolution
in circumstellar disks. The paper is organized as follows. In Section~\ref{two-D}, the
hydrodynamics equations in the thin-disk limit are reviewed. In Section~\ref{three-D},
we formulate the modifications made to the thin-disk model to improve the thermal balance calculations in the thin disk. In Section~\ref{comparison}, we compare the disk evolution in the 2+1D and 2D approaches.
The model caveats and future improvements are discussed in Section~\ref{caveats}. The main conclusions
are summarized in Section~\ref{conclude}. The Appendix presents details of the solution procedure
used to calculate the disk vertical structure. 

\section{Model equations in the thin-disk limit}
\label{two-D}
The equations of mass, momentum, and energy transport describing the dynamics of 
circumstellar disks in the thin-disk limit can be formulated as follows:
\begin{equation}
\label{cont}
\frac{{\partial \Sigma }}{{\partial t}} =  - \nabla  \cdot 
\left( \Sigma \bl{v} \right),  
\end{equation}
\begin{equation}
\label{mom}
\frac{\partial}{\partial t} \left( \Sigma \bl{v} \right) +  \nabla \cdot \left( \Sigma \bl{v}
\otimes \bl{v} \right) =   - \nabla {\cal P}  + \Sigma \, \bl{g} + \nabla \cdot \mathbf{\Pi}, 
\end{equation}
\begin{equation}
\label{energ}
\frac{\partial e}{\partial t} +\nabla \cdot \left( e \bl{v} \right) = -{\cal P} 
(\nabla \cdot \bl{v}) -\Lambda +\Gamma + 
\left(\nabla \bl{v}\right):\mathbf{\Pi}, 
\end{equation}
where $\Sigma$ is the mass surface density, $e$ is the internal energy per 
surface area, ${\cal P}$ is the vertically integrated gas pressure calculated 
via the ideal equation of state as ${\cal P}=(\gamma-1) e$ with $\gamma=7/5$, 
$\bl{v}=v_r \hat{\bl r}+ v_\phi \hat{\bl \phi}$ is the velocity in the
disk plane, and $\nabla=\hat{\bl r} \partial / \partial r + \hat{\bl \phi} r^{-1} 
\partial / \partial \phi $ is the gradient along the planar coordinates of the disk.
The gravitational acceleration in the disk plane, $\bl{g}=g_r \hat{\bl r} +g_\phi \hat{\bl \phi}$, 
takes into account self-gravity of the disk and the gravity of the central protostar when formed. 
Disk self-gravity is found  by solving for the Poisson integral
\begin{eqnarray} 
  \Phi(r,\phi) & = & - G \int_{r_{\rm sc}}^{r_{\rm out}} r^\prime dr^\prime 
                     \nonumber \\ 
      & &       \times \int_0^{2\pi} 
               \frac{\Sigma(r^\prime,\phi^\prime) d\phi^\prime} 
                    {\sqrt{{r^\prime}^2 + r^2 - 2 r r^\prime 
                       \cos(\phi^\prime - \phi) }}  \, ,
\label{gravpot}
\end{eqnarray} 
where  $r_{\rm sc}$ and $r_{\rm out}$ are the radial positions of the computational inner and
outer boundaries. This integral is calculated 
using a FFT technique which applies 
the 2D Fourier convolution theorem for polar coordinates 
and allows for the non-periodic boundary conditions in the $r$-direction  by effectively
doubling the computation domain in this coordinate direction and filling it with zero densities 
\citep[see][Sect.\ 2.8]{BT1987}.
Turbulent viscosity is taken into account via the viscous stress tensor 
$\mathbf{\Pi}$, the expression for which can be found in \citet{VB2010}.
The kinematic viscosity needed to calculate the viscous stress tensor is found 
adopting the Shakura and Sunyaev parameterization \citep{SS1973}, so that $\nu=\alpha c_{\rm s} h$,
where $c_{\rm s}=\sqrt{\gamma {\cal P}/\Sigma}$ is the sound speed and $h$ is the disk scale height.
The $\alpha$-parameter is linked to the inferred strength of the magneto-rotational instability 
(MRI) following the method that takes  the 
MRI active/inactive states into account as described in \citep{Bae2014}.

We use the following form for the cooling term $\Lambda$ in equation~(\ref{energ})
based on the analytical solution of the radiation transfer equations in the vertical  
direction \citep{Dong2016}:
\begin{equation}
\Lambda=\frac{4\tau_{\rm P} \sigma T_{\rm mp}^4 }{1+2\tau_{\rm P} + 
{3 \over 2}\tau_{\rm R}\tau_{\rm P}},
\label{cooling}
\end{equation}
where $\tau_{\rm R}=\kappa_{\rm R} \Sigma_{1/2}$ 
and $\tau_{\rm P}=\kappa_{\rm P} \Sigma_{1/2}$ the 
Rosseland and Planck optical depths to the disk midplane,  $\kappa_P$ and 
$\kappa_R$ the Planck and Rosseland mean opacities, and $\Sigma_{1/2}=\Sigma/2$ 
the gas surface density from the disk surface to the midplane. 
Similar forms of the disk cooling term were employed in other 1D axisymmetric and 
2D thin-disk simulations \citep[e.g.][]{Kley2008,Rice2009,VB2010,Zhu2012,Bae2014}  
adopted from analytic studies of the disk vertical structure by  
\citet{Hubeny1990} followed by slight modifications to the needs of numerical modeling by 
\citet{JG2003} 

The heating function $\Gamma$ is 
expressed by analogy to the cooling function as  \citep{Dong2016} 
\begin{equation}
\Gamma=\frac{4\tau_{\rm P} \sigma T_{\rm irr}^4 }{1+2\tau_{\rm P} + 
{3 \over 2}\tau_{\rm R}\tau_{\rm P}},
\label{heating}
\end{equation}
where $T_{\rm irr}$ is the irradiation temperature at the disk surface 
determined by the stellar and background black-body irradiation as
\begin{equation}
T_{\rm irr}^4=T_{\rm bg}^4+\frac{F_{\rm irr}(r)}{\sigma},
\label{Tirrad}
\end{equation}
where $T_{\rm bg}$ is the uniform background temperature (in our model set to the 
initial temperature of the natal cloud core)
and $F_{\rm irr}(r)$ is the radiation flux (energy per unit time per unit surface area) 
absorbed by the disk surface at radial distance 
$r$ from the central object. The latter quantity is calculated as 
\begin{equation}
F_{\rm irr}(r)= \frac{L_\ast}{4\pi r^2} \cos{\gamma_{\rm irr}},
\label{fluxF}
\end{equation}
where $\gamma_{\rm irr}$ is the incidence angle of 
radiation arriving at the disk surface at radial distance $r$. The incidence angle is calculated
using the disk surface curvature inferred from the radial profile of the  
disk vertical scale height \citep{VB2010}. The total stellar luminosity $L_\ast$ 
includes contributions from the accretion and photospheric luminosities.
Similar forms of the disk heating term were employed in other 1D axisymmetric and 
2D thin-disk simulations \citep[e.g.][]{Rice2009,Zhu2012,Bae2014}.

\section{Improving the thermal balance calculations: the 2+1D approach}
The numerical method described above is a fast and convenient
means for computing the disk evolution with high numerical resolution and for many
physical realizations \citep[e.g.][]{VB2010,Zhu2012}. However, this method is essentially
two-dimensional\footnote{though some modifications include a calculation of 
the disk vertical scale height and the incidence angle of stellar irradiation \citep{VB2010}.} 
and, as such, it lacks the information on the disk vertical structure. Some studies 
\citep[e.g.][]{Dong2016} assume a Gaussian or exponential vertical density profile,
but the same cannot be easily done for the vertical temperature distribution.
We therefore have developed a straightforward modification to this method, which
enables a calculation of the density and temperature distributions in the vertical
direction concurrently with the computations of the gas dynamics in disk plane.

\subsection{The 2+1D approach}
\label{three-D}
In this section, we formulate the modifications made to the thin-disk model
in order to improve the thermal balance calculations in the disk. 
These modifications also enable a reconstruction of the disk vertical structure, thus providing 
information on the disk volumetric density and temperature distributions.
The new equations of mass, momentum, and energy transport now read as follows:
\begin{equation}
\label{cont2}
\frac{{\partial \Sigma }}{{\partial t}} =  - \nabla  \cdot 
\left( \Sigma \bl{v} \right),  
\end{equation}
\begin{equation}
\label{mom2}
\frac{\partial}{\partial t} \left( \Sigma \bl{v} \right) +  \nabla \cdot \left( \Sigma \bl{v}
\otimes \bl{v} \right) =   - \nabla {\cal P}  + \Sigma \, \bl{g}
+ \nabla \cdot \mathbf{\Pi}, 
\end{equation}
\begin{equation}
\label{energ2}
\frac{\partial e}{\partial t} +\nabla \cdot \left( e \bl{v} \right) = -{\cal P} 
(\nabla \cdot \bl{v}) + \left(\nabla \bl{v}\right):\mathbf{\Pi}.
\end{equation}

While Equations~(\ref{cont2}) and (\ref{mom2}) remain 
essentially similar to their thin-disk counterparts (apart from the effect of stellar motion), 
Equation~(\ref{energ2}) now updates the internal energy only due to 
advection, viscous dissipation, and pressure work (adiabatic heating and cooling).
To take the disk heating by the stellar and background irradiation and the disk cooling 
due to its own infrared emission into account, we solve the moment equations 
describing the propagation of diffuse IR radiation in the vertical direction 
written in the Eddington approximation:
\begin{eqnarray}
\label{m1}
&&c_{\rm V} \dfrac{\partial T}{\partial t} = \kappa_{\rm P} c (E-aT^4) + S \label{m1} \\
&&\dfrac{\partial E}{\partial t} - \dfrac{\partial}{\partial z}
\left(\dfrac{c}{3\rho\kappa_{\rm R}} \dfrac{\partial E}{\partial z}\right) = -\rho \kappa_{\rm P} c(E-aT^4),
\label{m2}
\end{eqnarray}
where $E$ is the radiation energy density, $T$ the gas temperature, $\rho$ the gas volume density, 
$c_{V}$ the heat capacity of the gas, $c$ the speed of light,   $a$ the radiation constant, $z$ the vertical distance from the midplane, 
$\sigma$ the column density measured from the disk mid-plane, 
and $S$ the heating source (per unit mass) by the stellar and interstellar UV radiation.

Equations~(\ref{m1}) and (\ref{m2}) are complemented with the equation describing
the local vertical hydrostatic balance in the disk taking into account the gravity of the star
as well as the local self-gravity of the disk:
\begin{equation}
{R \over \mu \rho} \dfrac{d(\rho T)}{d z} =  -\dfrac{G M_\ast}{r^3}z - 4\pi G \sigma,
\label{static}
\end{equation}
where $M_\ast$ is the mass of the central star and $\mu=2.33$ the mean molecular weight. 
We note that $d \sigma = \rho dz$.  
We assume that the disk vertical columns at various positions in the disk 
do not influence each other, so that solving for Equations~(\ref{m1})-(\ref{static}) 
reduces to a series of 
1D problems for each grid zone on the ($r, \phi$) computational mesh.

\subsection{The solution procedure}
\label{solution}
Our solution method for Equations~(\ref{cont2})-(\ref{static}) 
consists of three steps. In the first, so-called source step, we update
the gas velocity and internal energy (per unit surface area) due to gravity,
viscosity and pressure work by solving for the following equations
\begin{equation}
\label{mom2_source}
\frac{\partial}{\partial t} \left( \Sigma \bl{v} \right)  
=   - \nabla {\cal P}  + \Sigma \, \bl{g} + \nabla \cdot \mathbf{\Pi}, 
\end{equation}
\begin{equation}
\label{energ2_source}
\frac{\partial e}{\partial t} = -{\cal P} 
(\nabla \cdot \bl{v}) + \left(\nabla \bl{v}\right):\mathbf{\Pi}.
\end{equation}

In the second, so-called "thermal" step, we compute the change 
in the disk gas temperature due to radiative cooling/heating and reconstruct the
disk vertical structure. To do that, we solve for 
the moment equations~(\ref{m1}) and (\ref{m2}) describing the propagation of diffuse IR radiation 
in the vertical direction complemented with equation~(\ref{static}) for the vertical hydrostatic balance.
We note that in Step~2 we use the gas temperatures that are partly updated in Step~1.
The detailed solution procedure for Step~2 is provided in the Appendix.

In the third, so-called transport step, we take advection of hydrodynamical quantities into account
by solving for the following equations:
\begin{equation}
\label{cont2_adv}
\frac{{\partial \Sigma }}{{\partial t}} + \nabla  \cdot 
\left( \Sigma \bl{v} \right) =0 ,  
\end{equation}
\begin{equation}
\label{mom2_adv}
\frac{\partial}{\partial t} \left( \Sigma \bl{v} \right) +  \nabla \cdot \left( \Sigma \bl{v}
\otimes \bl{v} \right) =  0
\end{equation}
\begin{equation}
\label{energ2_adv}
\frac{\partial e}{\partial t} +\nabla \cdot \left( e \bl{v} \right) = 0.
\end{equation} 
In this final step, we use the gas velocities and internal energies which 
are consequently updated during Steps~1 and~2.

To accomplish steps 1 and 3, we employ a combination of the finite difference and finite volume methods with a time-explicit
solution procedure similar in methodology to the ZEUS code \citep{SN1992}.
The advection in step 3 is treated using the third-order-accurate piecewise 
parabolic scheme of \citet{Colella1984}. A small amount of artificial viscosity 
is added to the code to smooth out shocks over two grid zones in both coordinate directions. 
The associated artificial viscosity torques
integrated over the disk area are negligible in comparison
with gravitational or turbulent viscosity torques. 

Equations~(\ref{mom2_source}), (\ref{energ2_source}), and
(\ref{cont2_adv})-(\ref{energ2_adv}) are discretized in polar coordinates ($r, \phi$) 
on a numerical grid with $512\times512$ grid zones. The radial points are
logarithmically spaced, while the azimuthal points are equidistant.  
The innermost grid point is located at
the position of the sink cell $r_{\rm sc}$ = 10 AU, and the size of the first
adjacent cell is 0.14~AU which corresponds to a radial resolution of $\Delta r=1.4$~AU 
at 100~AU. With this grid spacing, the Jeans
length  $R_{\rm J}= \langle v^2 \rangle/\pi G \Sigma$, where $ \langle v^2 \rangle$ is the velocity dispersion
in the disk plane \citep{Vor2013}, is resolved by roughly 10--20 grid zones in each coordinate 
direction to radial distance up to 500~AU, thus fulfilling the Truelove criterion \citep{Truelove1998}.

Equations~(\ref{m1})--(\ref{static}) are solved on an adaptive, non-equidistant grid
with 32 grid points. The finest grid spacing is usually in the disk atmosphere where 
the largest density gradients are found.
The inner and outer boundaries on the polar grid ($r,\phi$) allow for material to freely flow out
from the active computational domain, but prevent any material to flow in. 
In the vertical direction, we assume a reflecting boundary in the disk midplane 
and a constant gas volume density of $10^3$~cm$^{-3}$ at the disk upper edge.

\begin{table*}
\renewcommand{\arraystretch}{1.2}
\center
\caption{Model core parameters}
\label{table1}
%\vspace{3 pt}
\begin{tabular}{ccccccc }
\hline\hline
Model & $M_{\rm core}$ & $\beta$  & $\Omega_0$  & $r_{\rm 0}$ & $\Sigma_0$  
& $R_{\rm out}$  \\
 & ($M_\odot$) & ($\%$) & (km~s$^{-1}$~pc$^{-1}$) &  (AU) & (g~cm$^{-2}$) & (pc)  \\
\hline
A & 1.38 & 0.5 &  1.53  & 2057 & $9.05\times 10^{-2}$ & 0.06  \\
B & 1.38 & 0.27 & 1.13  & 2057 & $9.05\times 10^{-2}$ & 0.06  \\
\hline
\end{tabular}
\end{table*}

\subsection{Initial conditions}
\label{initial}
Our numerical simulations start from a pre-stellar core with the radial profiles
of column density $\Sigma$ and angular velocity $\Omega$ described as follows:
\begin{eqnarray}
\Sigma(r) & = & {r_0 \Sigma_0 \over \sqrt{r^2+r_0^2}}\:, \\
\Omega(r) & = &2\Omega_0 \left( {r_0\over r}\right)^2 \left[\sqrt{1+\left({r\over r_0}\right)^2
} -1\right],
\label{ic}
\end{eqnarray}
where $\Sigma_0$ and $\Omega_0$ are the gas surface density and angular velocity 
at the center of the core. These profiles have a small near-uniform
central region of size $r_0$ and then transition to an $r^{-1}$ profile;
they are representative of a wide class of observations and theoretical models
\citep{Andre93,Dapp09}. 

Our iterative solution procedure for Equations~(\ref{m1})--(\ref{static}) (see the Appendix),
requires us to make an initial guess for the vertical structure of the core. 
We assume a constant gas temperature
of 15~K and a Gaussian distribution of the gas volume density  of the form 
\begin{equation}
\rho(z)=\rho_0 e^{ -\left( z/h  \right)^2},
\end{equation}
where $\rho_0=\Sigma/(h\sqrt{\pi})$.
We note, however, that these initial conditions is a mere initialization requirement and 
the vertical structure of the core quickly attains the form determined by the combined action of
the external heating, radiative cooling,  pressure gradients, and self-gravity of the core. 

We have considered several model cores, but in this paper, for the sake of conciseness, we present 
only two. The initial parameters of these models are
shown in Table~\ref{table1},  where $M_{\rm c}$ is the initial core mass, $\beta$ 
the ratio of rotational energy to the magnitude of gravitational potential energy,
and $r_{\rm out}$ the initial radius of the core. The two models are different in the amount of initial
rotation, as manifested by distinct $\beta$ and $\Omega_0$.  
The initial parameters are chosen so that the cores are gravitationally unstable and
begin to collapse at the onset of numerical simulations. 
We monitor the gas surface density in the sink cell, 
and when its value exceeds a critical one for the transition from isothermal 
to adiabatic evolution, we introduce a central point-mass star. In the subsequent
evolution,  90\% of the gas that crosses the inner boundary is assumed to land onto
the growing star. The other 10\% of the accreted
gas is assumed to be carried away with protostellar jets  
The simulations continue into the embedded phase of star
formation, during which a protostellar disk is formed. 
The simulations are terminated in the T Tauri phase when nearly all material of 
the parental core has accreted onto the resulting star-plus-disk system.

\begin{figure*}
\begin{centering}
\includegraphics[scale=0.65]{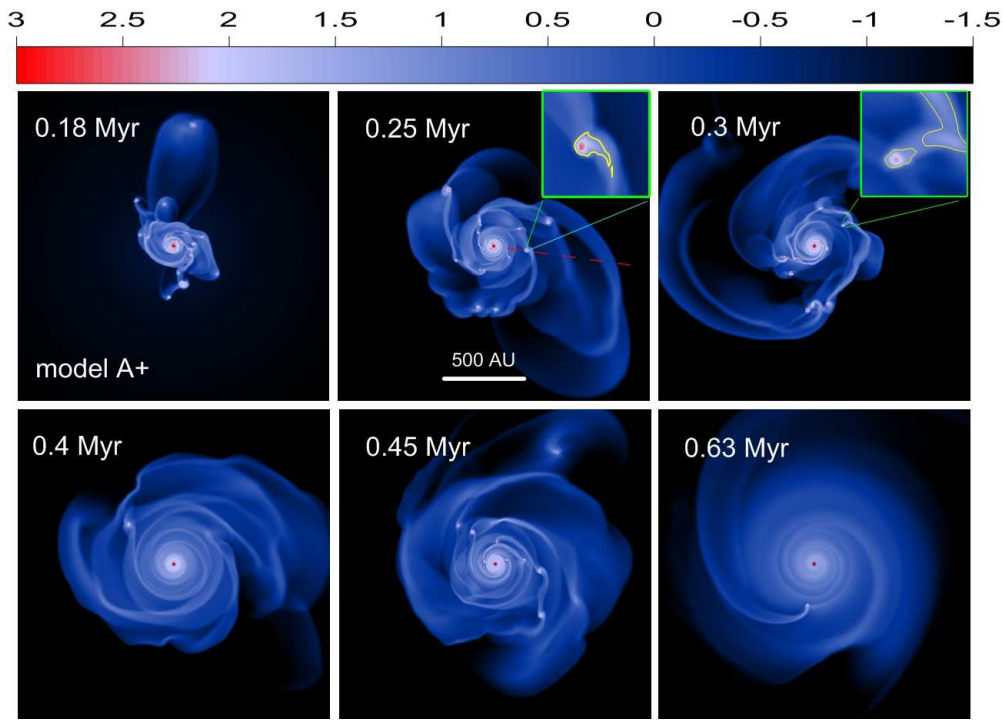} 
\par\end{centering}
\protect\protect\caption{\label{fig1}Gas surface density distributions in model~A+ shown at several
time instances after the onset of numerical simulations (the disk forms at $t=0.1$~Myr). The inserts zoom in onto several fragments in the disk. The
yellow contour lines delineate regions with the Toomre $Q$-parameter smaller than unity.  The red dashed
line is a radial cut through the disk used later to show the vertical volume density and temperature
distributions. The scale bar is in $\log$~g~cm$^{-2}$.}
\end{figure*}
\begin{figure*}
\begin{centering}
\includegraphics[scale=0.65]{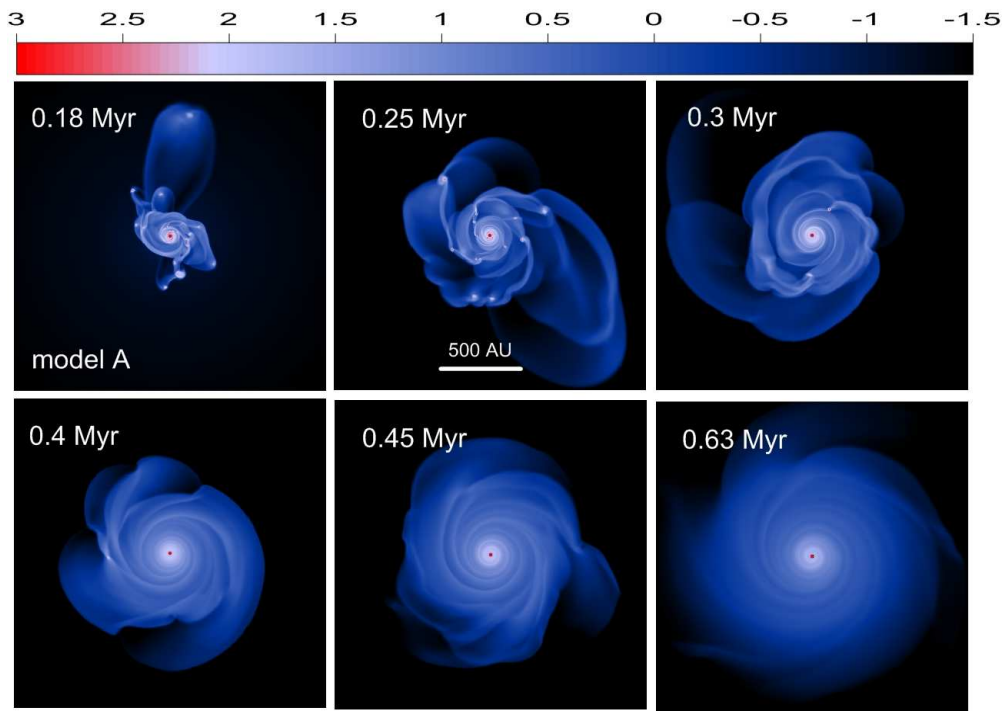}
\par\end{centering}
\protect\protect\caption{\label{fig2}Similar to Figure~\ref{fig1}, but for model~A.}
\end{figure*}

\section{Comparison of 2+1D and 2D approaches}
\label{comparison}
In this section, we compare the properties of circumstellar disks obtained using the original 
2D thin-disk model and the improved 2+1D model 
with the purpose to understand how an improved calculation of the thermal balance in the disk
can effect its dynamical evolution.
First, we compute the disk evolution using the 2+1D models starting from the collapse 
of pre-stellar cores
and ending in the T Tauri stage of disk evolution when the disk age exceeds 0.5~Myr. 
Then, we compute the disk evolution in the 2D model starting 
from a certain time instance, usually, in the Class~I phase, using the current values
of $\Sigma$, $e$, and $\bl{v}$ taken from the 2+1D simulations.
This allows us to directly determine the effect of different thermodynamical schemes 
on the disk dynamical evolution and avoid the possible dynamical influence of 
the early collapse phase.  In the following text, the models computed using the 2+1D
approach are distinguished with the "plus" sign.

\subsection{Models~A+ and A}

Figure~\ref{fig1} presents the disk evolution in model~A+ computed using the 2+1D approach. 
Shown are the gas surface density snapshots in the inner $2000\times2000$~AU$^{2}$ box 
taken at several times elapsed since the onset of numerical simulations. 
The entire numerical domain is 
about 10 times larger and includes the infalling envelope. Evidently, the disk is 
strongly gravitationally unstable and multiple fragments are seen forming in the disk's 
middle and outer regions. The Toomre Q-parameter,  $Q=c_{\rm s} \Omega/\pi G \Sigma$, 
is smaller than unity in and around the fragments,
as is demonstrated by the yellow contour lines in the inserts of Figure~\ref{fig1}.
The masses of these fragments range from about that of a Jupiter to the upper-mass brown dwarfs.

This behaviour is typical for massive, non-magnetized or weakly-magnetized disks in the embedded
phase of evolution (especially, in the Class I phase), 
where gravitational instability is fueled by continuous mass loading from
the infalling parental cloud \citep{VB2005,VB2015,Kratter2008,Tsukamoto2013,Meyer2017}. 
In model~A+, the embedded phase ends around $t=0.22$~Myr, but fragmentation continues 
into a later phase because the disk is massive with the disk-to-star 
mass ratio $\sim0.2$ at the end of the embedded phase. Many fragments do not live long 
and migrate into the star, causing FU-Orionis-type luminosity outbursts 
\citep{VB2005,VB2015,Machida2011}. Others are dispersed by tidal torques \citep{Vor2011,Zhu2012} or ejected from the disk via multi-body 
gravitational interactions \citep{Stamatellos2009,BV2012,Vor2016}.
However, some fragments may survive, contract to planetary sizes, and form planets or
brown dwarfs at various distances from the star \citep{Boss1998,Nayakshin2010,Nayakshin2011, 
Boley2010,Vorobyov2013,Stamatellos2015a,Stamatellos2015b,Galvagni2014}.
Disk fragmentation, therefore,  can be an important channel for the formation of planets and brown dwarfs,
either as companions to the host star or as freely floating objects.

Figure~\ref{fig2} presents the disk evolution in model~A computed using the 2D approach. The gas surface
density maps at the same evolutionary times as in model~A+ are shown in the inner $2000\times2000$~AU$^2$
box. The 2D simulations start from $t=0.25$~Myr, explaining why the disk appearance at $t\le0.25$~Myr
is identical in models~A+ and A. At later times, however, notable differences in the disk evolution between model~A+ and model~A arise. While the disk in model~A+ is strongly unstable and 
often harbours multiple fragments, the disk in model~A
experiences only occasional fragmentation after $t=0.25$~Myr. 

This difference is further illustrated
in Figure~3 showing the number of fragments in the disk $N_{\rm frag}$ at a certain time instance 
as a function of time elapsed since the
onset of numerical simulations. Two conditions were used to identify the fragments in the
disk~\citep[see][for detail]{Vor2013}:\\
1) they must be pressure supported,
with a negative pressure gradient with respect to the
center of the fragment; and 2) they must be kept together
by gravity, with the potential well being deepest at the center of
the fragment.   Evidently, the number of fragments in the disk at a given time instance is greater
in model~A+ than in model~A. In addition, the duration of the disk fragmentation stage 
(defined as the time span during which fragments are present in the disk) 
is longer in model~A+. Both evidence suggest that the disk in model~A+ is more gravitationally
unstable. 

\begin{figure}
\begin{centering}
\resizebox{\hsize}{!}{\includegraphics{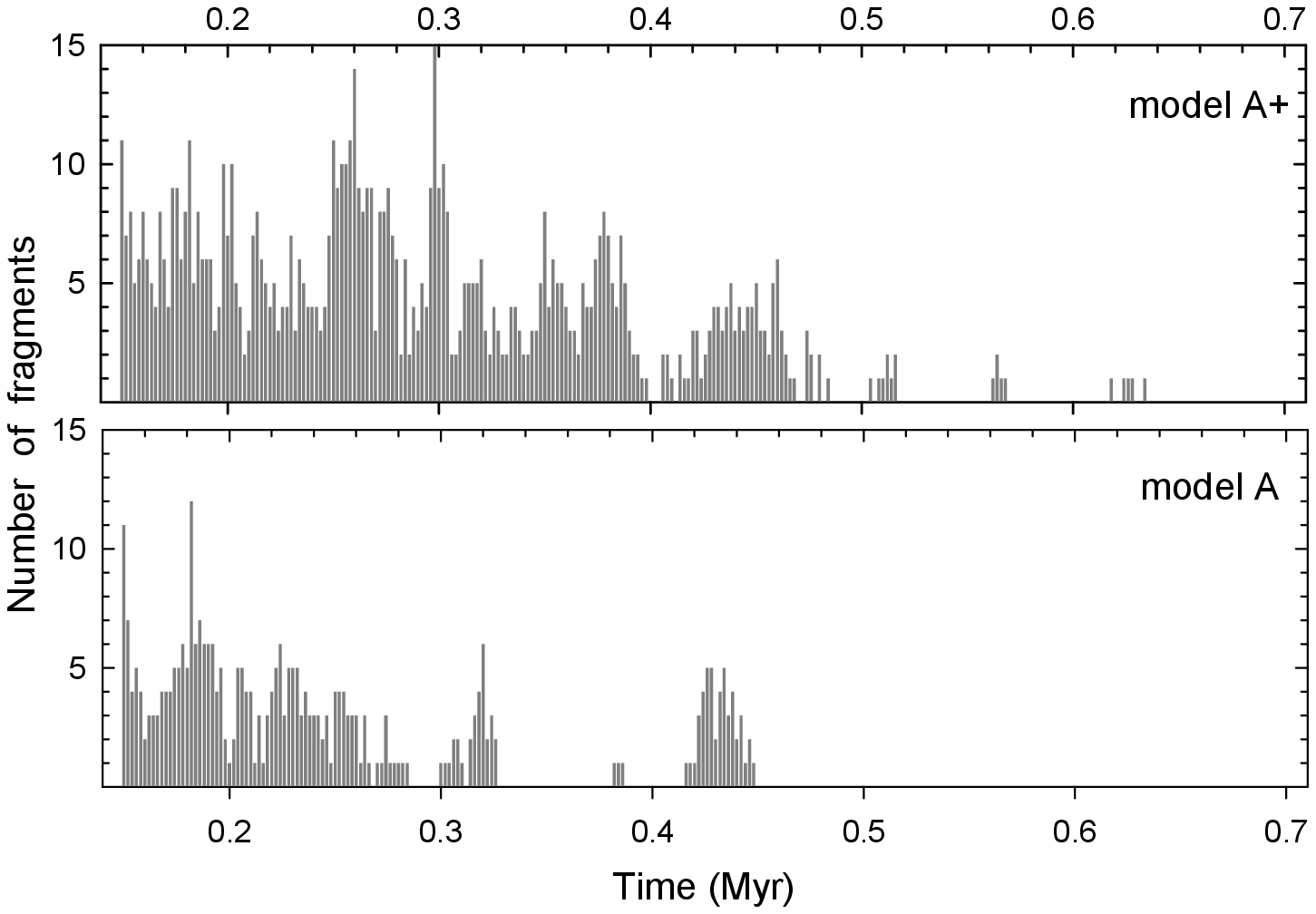}}
\par\end{centering}
\centering{}\protect\protect\protect\caption{\label{fig3} Number of fragments 
in the disk at a certain time instance as a function of time elapsed since the
onset of numerical simulations in model~A+ (top) and model~A (bottom).
The disk forms at $t=0.1$~Myr.}
\end{figure}

This difference in the strength of gravitational instability can in principle be caused 
by distinct disk and stellar masses in models~A+ and A. It is known that systems
with a higher disk-to-star mass ratio are characterized by stronger gravitational instability
because they have on average lower angular velocities and/or higher surface densities, and are thus characterized by lower $Q$-parameters. However,
we found that model~A+ has a slightly lower disk mass and a slightly higher stellar
mass than model~A, {meaning that
the mass transport rate through the disk in this model is systematically higher than in model~A.
The higher mass transport rates in model~A+ are caused by systematically higher gravitational torques,
as can be expected for disks with stronger gravitational instability.}

The stronger gravitational instability in model~A+ as compared 
to that in model~A is not related to the disk or stellar masses. It may then be related
by differences in the disk thermal structure arising from distinct approaches to calculating
the disk cooling and heating in 2+1D and 2D approaches. To check if this is indeed the case, 
we plot in Figure~\ref{fig6} the radial gas temperatures profiles in model~A+ (red lines) 
and model~A (black lines),
calculated as $T_{\rm hydro}=e(\gamma-1)\mu/\Sigma {\cal R}$ 
for every cell on the polar grid ($r,\phi$) 
and then arithmetically averaged over the azimuth. We note that in the 2+1D approach, $T_{\rm hydro}$ represents the gas temperature mass-weighted over the vertical column of gas 
in each cell ($r,\phi$) (see Equation~\ref{energy2D}). 
{For model~A+, we also plot the temperature in the disk midplane $T_{\rm mp}$ (blue lines).} 
Four time instances elapsed since the onset of both 2+1D and 2D simulations are shown.

Evidently, notable differences in the radial gas temperature distributions develop with time 
in models~A+ and A. The disk in model~A+ is systematically colder than in model~A, {no matter
what temperature in model~A+ is considered: the hydrodynamic ($T_{\rm hydro}$) or the midplane 
one ($T_{\rm mp}$)}. On the contrary,
the inner envelope is warmer in model~A+ than in model~A. 
{These differences can be understood from the following analysis. Let us first consider 
the 2D case and for a moment  neglect the heating sources due to viscosity and compressional heating
due to $P dV$ work. 
In the steady-state case, the temperature in the disk and envelope will be controlled by a balance between radiative cooling
$\Lambda$ (equation~\ref{cooling}) and irradiation and background heating $\Gamma$ (equation~\ref{heating}),
so that the midplane temperature can be written as
\begin{equation}
T_{\rm mp}^4 = T_{\rm bg}^4 + {F_{\rm irr}\over \sigma}.
\label{T2Dp1}
\end{equation}
This steady-state temperature is plotted in Figure~\ref{fig6} with the dashed black lines. Evidently,
it is very close to the actual temperature in model~A everywhere except the very inner parts of the
disk where viscous and compressional heating become substantial and the midplane temperature
rises above the analytically predicted values. 

\begin{figure}
\begin{centering}
\resizebox{\hsize}{!}{\includegraphics{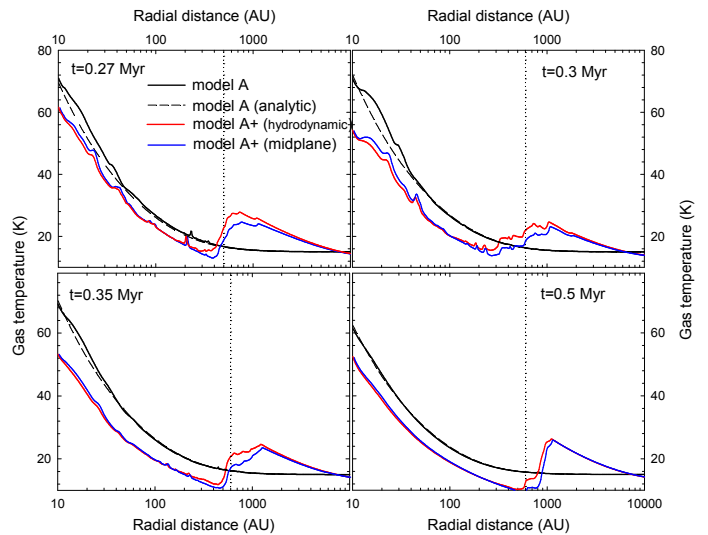}}
\par\end{centering}
\centering{}\protect\protect\protect\caption{\label{fig6} Azimuthally averaged radial profiles of the
gas temperature in model~A+ (the red and blue lines) and model~A (the black solid lines) taken at several
times since onset of numerical simulations. {In particular, the red lines represent the gas temperature
mass-weighted over the vertical column of gas, while the blue lines are the gas temperature in the disk
midplane. The dashed solid lines are the gas temperatures in model~A derived for the steady-state
case neglecting compressional and viscous heating. }
The vertical
dotted lines mark the radial positions where the disk joins the infalling envelope.}
\end{figure}

In the2+1D case, the midplane temperature can be expressed in the following form if the medium is optically
thick to ultraviolet and  infrared radiation
\begin{equation}
T_{\rm mp}^4 = {F_{\rm irr} \over 2 \sigma}.
\label{T2D}
\end{equation}
This expression follows from the assumption that the incident UV flux absorbed by dust in the disk 
is radiated away isotropically to the upper and lower
hemisphere \citep[see, e.g., equation 12a in][]{Chiang1997}. 

A comparison of Equations~(\ref{T2Dp1}) and (\ref{T2D}) in the regions where the input from the 
background irradiation is much smaller than from the stellar one (i.e., in the disk), demonstrates
that the midplane temperature in model~A+ is expected to be a factor of $2^{1/4}\approx1.2$ 
smaller than that of model~A. A similar difference between the gas temperatures in models~A+ and A
is also seen in Figure~\ref{fig6}. We note that the stellar luminosity in our models exhibits short-term
variations caused by the time-variable protostellar accretion \citep[e.g.][]{VB2015}. As a result,
the disk thermal state may take some time to adjust to the time-variable stellar flux, 
meaning that the midplane temperature may not exactly coincide with the analytical values derived 
in the steady-state limit of constant stellar luminosity. 
}

{In the optically thin (to UV radiation) limit, the midplane temperature in the 2+1D model can be
written as
\begin{equation}
T_{\rm mp}^4 = {\kappa_{\rm F} \over \kappa_{\rm R}} {F_{\rm irr} \over 2 \sigma},
\label{thin2Dp1}
\end{equation}
where $\kappa_{F}$ is the dust opacity in the UV band. Evidently, the midplane temperature in 
the optically thin
limit is higher than in the optically thick one by a factor of $(\kappa_{\rm F}/\kappa_{\rm R})^{1/4}$.
The optically thin limit is expected to take place in the envelope and this explains an increase
in the gas temperature in model~A+ at distances $> 1000$~AU. In the 2D case, the gas temperature 
in the envelope is controlled by the background irradiation with $T_{\rm bg}=15$~K. At the same 
time, Equation~(\ref{thin2Dp1}) for $F_{\rm irr}= D T_{\rm ISM}^4$ (only the interstellar component,
see Appendix~A) 
yields the midplane temperature $\approx 14$~K for $\kappa_{\rm F}/\kappa_{\rm R}=10^3$, 
demonstrating that the midplane temperatures in both models converge to a similar value in the outer envelope.
}

The consequences of this distinct radial temperature distribution is twofold.
First, a lower gas temperature makes the disk in model~A+ more gravitationally unstable, 
as was already noted above. Second, higher infall rates from the collapsing envelope 
onto the disk $\dot{M}_{\rm infall}\sim c_{\rm
s}^3/G$, as implied by a higher gas temperature in the inner envelope, 
drive the disk in model~A+ faster to the critical point at which the disk becomes unstable 
to fragmentation, thus creating more fragments in the disk. In other words, the characteristic time
of mass infall onto the disk (or the disk mass replenishment) $M_{\rm disk}/\dot{M}_{\rm infall}$ is shorter in model~A+.

\begin{figure}
\begin{centering}
\resizebox{0.85\hsize}{!}{\includegraphics{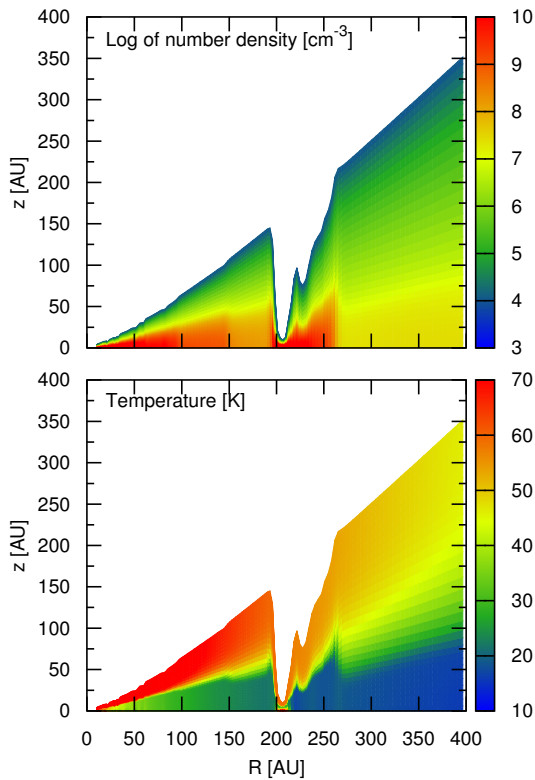}}
\par\end{centering}
\centering{}\protect\protect\protect\caption{\label{fig7} 
The distributions of gas volume density (top) and temperature
(bottom) in the $(r,z)$ plane  taken in model~A+ at $t=0.25$~Myr through the radial cut
with a position angle of $351^\circ$  (shown in Figure~\ref{fig1} with
the red dashed line).}
\end{figure}

In Figure~\ref{fig7} we present the two-dimensional distributions  of
the gas volume density and temperature in the 
($r,z$) plane obtained in model~A+ at $t=0.25$~Myr
by taking a vertical cut with a position angle of $351^\circ$  (shown in
Figure~\ref{fig1} with the red dashed line). The cut is chosen so that
it passes through an inner spiral arm at $r\approx 48$~AU and a dense
fragment with mass $\approx 7.8~M_{\rm Jup}$  at $r\approx 207$~AU
(shown in the insert of the upper-middle panel in Figure~\ref{fig1}).
Clearly, the gas volume density is
highest in the midplane and drops down toward the disk atmosphere. At the
same time, the temperature is higher in the upper layers and is minimal 
in the disk midplane everywhere in the disk, except for the fragments. 
This means that the temperature in all parts of the disk, except for the fragments, 
is controlled by the external radiation (stellar and background). 
In the fragments, however,
the gas is intensively heated by compressional and viscous heating, while the radiative
diffusion is not fast enough to cool the inner, optically thick parts of the fragment
down to low temperatures observed around the fragment. 
We also note that the fragment at $r=207$~AU is gravitationally bound, which results 
in a very low disk vertical height at and around this point. The signature of disk self-gravity is also
seen at $r=230-250$~AU covering the spiral arc in which the fragment resides and
where the disk vertical height is also significantly reduced as
compared with the surroundings. It is therefore possible that fragments may be deeply hidden
in the disk, which would  make their detection via observations of the near-infrared light
scattered off the disk surface quite difficult, as was already noted in \citet{Dong2016}. Fully three-dimensional
simulations are needed to determine how far from the midplane the fragments can be scattered via gravitational
interaction with other fragments and spiral arms.

\begin{figure}
%\begin{figure*}
\begin{centering}
\resizebox{\hsize}{!}{\includegraphics{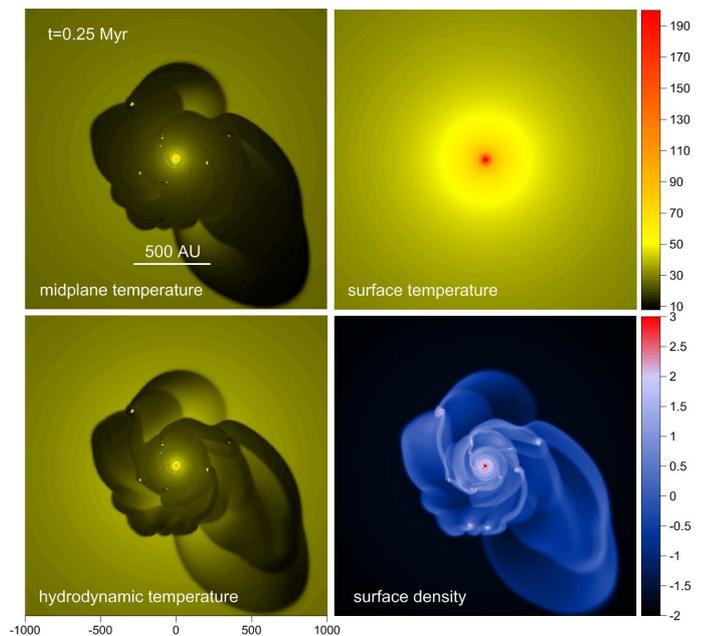}}
\par\end{centering}
\centering{}\protect\protect\protect\caption{\label{fig7a} 
Temperature and gas surface density distributions in model~A+. Various definitions of temperature
as described in the text are shown. }
%\end{figure*}
\end{figure}

The 2D temperature distributions in the disk midplane and at the surface of the disk are 
shown for model~A+ in the upper-left and upper-right panels of Figure~\ref{fig7a}. 
The bottom-left panel presents the gas hydrodynamic temperature defined as 
$T_{\rm hydro} = e (\gamma-1)\mu /(\Sigma {\cal R})$,  in a similar fashion 
as the midplane temperature in the 2D approach (see Equation~\ref{cooling}).
The gas surface density distribution is also shown in the bottom-right panel for convenience. 
All distributions are plotted at $t=0.25$~Myr.

The hydrodynamic and midplane temperatures reflect the 
structure of the disk --
both are rather low in the outer disk regions and dense spiral arms, 
but grow in the inner disk thanks mainly to increased stellar irradiation. 
Dense fragments heated by compressional and viscous heating stand out as bright spots
in the disk. 
{The hydrodynamic temperature $T_{\rm hydro}$ appears to be higher than the midplane temperature $T_{\rm mp}$ in the disk outer
regions and in the inner envelope. This trend can also be seen in Figure~\ref{fig6} and is 
explained by the passively heated (by stellar and interstellar irradiation) nature of the
outer disk and envelope -- the upper gas layers are always warmer than the midplane. As a result,
$T_{\rm hydro}$ -- a vertical average over the gas column -- becomes higher than $T_{\rm mp}$. In the inner disk regions, where efficient 
viscous and compressional heating operate in the disk midplane, the situation may reverse and 
$T_{\rm mp}$ may become higher than  $T_{\rm hydro}$. This trend is also evident in Figure~\ref{fig6}.}
The surface temperature smoothly decreases with radial distance 
from the star. The lack of azimuthal variations is a mere consequence of the adopted 
scheme for calculating the incidence angle of radiation onto the disk surface, 
which uses an azimuthally averaged disk scale height to calculate $\gamma_{\rm irr}$. 
We note that the surface temperature is higher than the midplane temperature 
almost everywhere in the disk, except for the fragments as discussed in more detail below.

\begin{figure}
\begin{centering}
\resizebox{\hsize}{!}{\includegraphics{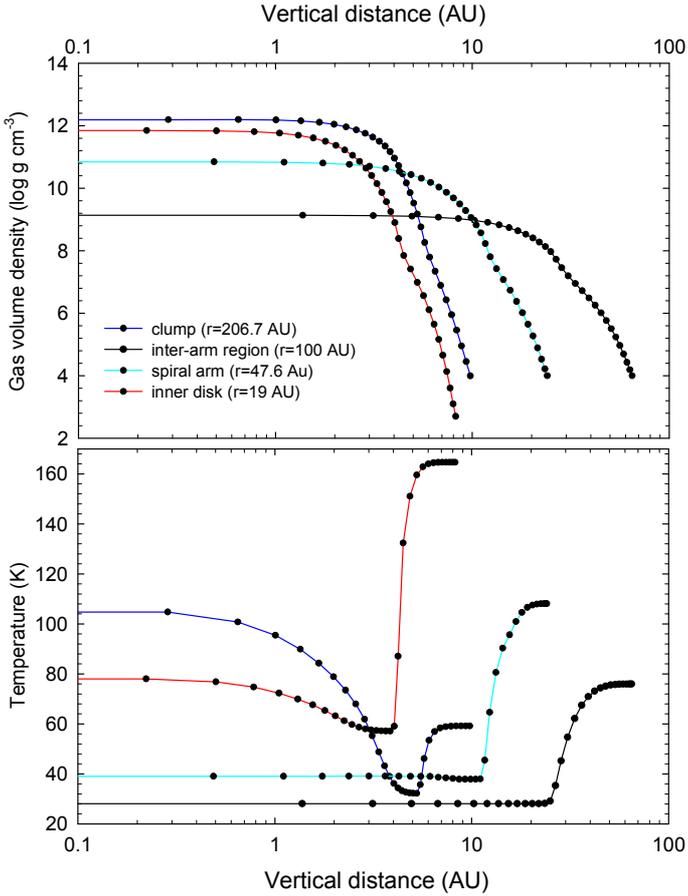}}
\par\end{centering}
\centering{}\protect\protect\protect\caption{\label{fig8} One-dimensional vertical 
cuts taken at different positions of the disk (as indicated in the legend) 
and showing the vertical gas volume density (top)
and temperature (bottom) distributions.}
\end{figure}

Finally, in Figure~\ref{fig8} we show the one-dimensional vertical gas volume density 
and temperature profiles taken at several positions in the radial cut 
(the red dashed line in Figure~\ref{fig1}). 
These positions were chosen so
as to show the vertical distributions in various sub-structures that may be present in the disk, such as spiral arms and fragments. More specifically, the blue lines represent the vertical profiles taken
through the fragment shown in the insert of the upper-middle panel in Figure~\ref{fig1}, 
while the cyan lines are taken through the spiral arm. The
black lines represent the vertical profiles in a regular inter-arm disk region and 
{the red lines are the vertical profiles in the inner disk.}
{We note that we apply an adaptive grid in the vertical direction that enables an adequate 
numerical resolution in the disk regions with strong volume density gradients and also 
in the disk atmosphere.} 

Evidently, the 
fragment is characterized by the highest volume density amounting to $10^{12}$~cm$^{-3}$. 
At the same time, 
the fragment has the most compact structure with a vertical size of
just 16~AU, notwithstanding the fact that it is actually located quite far away from the star. 
The midplane size of this fragment is $\approx 30$~AU 
so that the clump has an ellipsoidal shape with the semi-axis ratio of 
$a_{\rm z}:a_{\rm r}=1:2$. This scaling is in agreement with the ratio of rotational-to-thermal 
energy of 44\%, implying that the fragment has a substantial rotational 
support against gravity (in case of zero rotation, we would have expected a near-spherical shape). 
The fragment is also characterized by a peculiar vertical temperature distribution:
its midplane temperature is higher than the temperature in its atmosphere. {There is also 
a depression in the temperature profile in between the disk midplane and atmosphere. 
 A similar, non-regular vertical temperature distribution can also found in the inner disk regions
(the red line). In this case, however, the temperature in the atmosphere is somewhat higher
than in the disk midplane.} The non-regular temperature profiles are a direct 
consequence of the compressional and viscous heating operating in the optically thick 
interiors of the fragment and in the inner disk regions. 
These heating sources are more efficient than heating due to 
stellar irradiation, the latter being effectively absorbed by the atmosphere of the fragment.
This is an important phenomenon, which means that radiation transfer codes that neglect 
hydrodynamic heating sources (such as RADMC-3D) cannot accurately reproduce the temperature 
structure in the fragments formed via disk gravitational fragmentation and in the inner disk regions
\citep[see also][]{Dong2016}. The other two disk elements,
the spiral arm and the inter-arm region have a vertical temperature distribution typical for passively
irradiated disks, though the spiral arm demonstrates a mild increase of the 
gas temperature towards the midplane, caused probably by mild shock and viscous heating.

\subsection{Models~B+ and B}
Model~B+ is distinct from model~A+ in the amount of rotation initially present in the parental collapsing
core (see Table~\ref{table1}). More specifically, model~B+ has a lower ratio of rotational to gravitational
energy $\beta$, which implies that model~B+ forms a less massive disk than model~A+. 
Following the same procedure as was discussed in the beginning of Section~\ref{comparison}, 
we first computed the disk
evolution in model~B+ and then we computed model~B starting from a time instance 
$t=0.215$~Myr, which corresponds to the Class I phase of disk evolution.

\begin{figure}
\begin{centering}
\resizebox{\hsize}{!}{\includegraphics{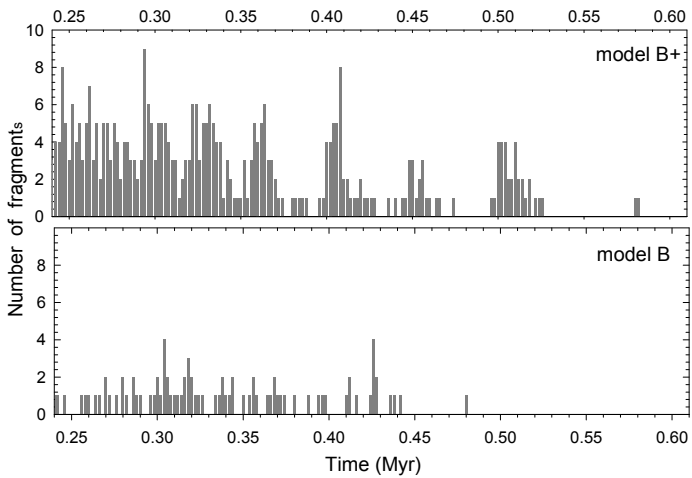}}
\par\end{centering}
\centering{}\protect\protect\protect\caption{\label{fig9} Number of fragments 
in the disk at a certain time instance as a function of time elapsed since the onset 
of numerical simulations 
in model~B+ (top) and model~B (bottom). The disk forms at $t=0.138$~Myr.}
\end{figure}

A comparison of model~B+ and model~B has revealed trends that are essentially similar to those
discussed in details for model~A+ and model~A. The disk in model~B+ is more gravitationally unstable,
more prone to fragmentation, and, at the same time, is less massive than in model~B. This apparent
paradox is explained by a systematically colder disk temperatures in model~B+ as compared to model~B.
For the sake of conciseness, we do not perform an in-depth analysis of models B+ and B 
in this section, but simply present in  Figure~\ref{fig9} the number of fragments in the disk $N_{\rm
frag}$ at a given time instance as a function of time elapsed since the onset of numerical simulations. Clearly, $N_{\rm frag}$ is larger in model~B+ than in model~B. The number of fragments in model~B+ 
can become as high as 9 in the early evolutionary phase, whereas in model~B the number of 
fragments never exceeds four.
The duration of the disk fragmentation phase is also longer in Model~B+, which implies a higher
detection likelihood of disk fragmentation.

\section{Model caveats and further improvements}
\label{caveats}
In this section, we discuss the model caveats and further improvements of our 2+1D model.

{\it Disk self-gravity}. When reconstructing the disk vertical structure,
a local value for the gas column density $\sigma$ was used in the last term of 
Equation~(\ref{static}). Under this approximation, the vertical gas columns
do not affect each other when solving the equation of the vertical
hydrostatic balance. This may affect the disk vertical structure in 
the regions dominated by gravity. 
For instance, a sharp drop in the disk height seen around the fragment 
in the upper panel of Figure~\ref{fig7} may be an artifact of the adopted
approximation.  In the future, the model needs to be improved 
by taking the non-locality of disk gravity into account.

{\it Vertical motions.} Another assumption behind the presented model
is the neglection of vertical motions. Namely, we assume that the disk
attains a vertical hydrostatic equilibrium on time scales much shorter
than the dynamical timescale. This approximation is justified if the dynamical timescale 
is longer than any other pertinent timescale, such as the timescales for diffusion of radiation,
 propagation of sound waves, stellar heating, which was shown to be usually the case for circumstellar
 disks in \citet{Vorobyov:2014}. With the current approach, we cannot model interesting effects, 
such as vertical oscillations and surface waves. This is, however, the price that we are willing 
pay for the ability to follow the disk evolution on much longer timescales than is currently 
possible with the
full 3D numerical hydrodynamics simulations. 

{\it Realistic equation of state}. In the current paper, we did not take into account 
that the heat capacity $c_{\rm v}$ and the adiabatic index $\gamma$ of gas depend on
the temperature since the rotation levels of molecular hydrogen can be
excited in the considered temperature ranges. We did not include this
effect since it requires a more complicated procedure for the 
solution of the radiation transfer equations. We plan to do this in follow-up papers.

There are other modifications to our radiative transfer
model which can be implemented in the future. Currently, to 
calculate heating due to the stellar UV radiation, we evaluate the
incidence angle of stellar radiation $\gamma_{\rm irr}$  (see Equation~\ref{UV-RT}) in each 
grid cell based on the local gradients of the disk scale height and then average the resulting
values over the azimuth \citep{VB2010}.  This procedure provides the radially varying
UV heating due to the global disk flaring, but does not allow us
to model properly the local effects, such as shadows in the disk caused by spiral arms. 
Our attempt to calculate the azimuthally varying $\gamma_{\rm irr}$, including the effect of the shadows,
resulted in overheating of the disk surfaces directly exposed to stellar radiation.
Allowing for the diffusion of the thermal IR emission  in the equatorial plane 
may alleviate this problem.
Finally, we now work on implementing the FARGO mechanism for advection of hydrodynamical quantities
\citep{FARGO},
which will allow us to ease the restrictive time step limitations and  move the sink cell boundary 
closer to the star, hopefully, to the sub-AU region.

\section{Conclusions}
\label{conclude}
In this paper, we have improved 
the frequently used thin-disk models of circumstellar disk evolution and presented the method
that includes a better 
calculation of the disk thermal balance and a reconstruction of the disk vertical 
structure. Our method is based on the solution of
the hydrodynamics equations describing the gas dynamics in the plane of the disk, 
complemented with solution of the radiation transfer and hydrostatic balance equations describing the
disk vertical structure. We performed a detailed comparison of 
this so-called 2+1D method with the purely 2D thin-disk models of disk evolution. 
Our findings can be summarized as follows.
\begin{itemize}
\item Improved 2+1D models yield systematically colder disks, while 
the infalling parental clouds in the embedded evolutionary phase are warmer. Both
effects act to increase the strength of disk gravitational instability in 2+1D models as compared 
to purely 2D simulations.

\item Disk gravitational fragmentation is more efficient and the duration of the disk fragmentation phase is longer in 2+1D models, which implies an increased likelihood for detecting 
disk fragmentation in protostellar disks.

\item The outer disk regions
are mostly characterized by a positive vertical temperature gradient, typical 
for so-called passive circumstellar disks heated mainly by stellar and background irradiation. 
{The inner disk regions usually have a more complex, non-regular vertical temperature distribution
having local peaks in the midplane and in the atmosphere separated by a mild depression. The 
temperature increase in the midplane is caused by efficient viscous and compressional heating operating
in the inner disk.}

\item
Fragments forming in the disk via gravitational fragmentation are also characterized by {a 
double-peaked vertical temperature profile, but unlike the inner disk regions, 
the center of the fragment is warmer than its atmosphere}. This means that 
the hydrodynamical heating sources (compression, viscosity) are more efficient than stellar 
irradiation heating, implying that radiation transfer codes that neglect the hydrodynamical 
heating sources cannot accurately compute the temperature
in the fragment interiors \citep[see also][]{Dong2016}.

\item Fragments are significantly more compact than the surrounding disk, which could make their detection
with the scattered light techniques difficult unless significant excursions away from the disk midplane are feasible.

\end{itemize}

A detailed procedure for solving the radiation transfer and hydrostatic balance equations 
in the vertical direction is presented in the Appendix.
The improved 2+1D method yield a full 3D structure of the disk, namely its volumetric density and temperature
distributions, with a modest increase in the net computational time with respect to purely 
2D models. {It also applies an adaptive mesh in the vertical direction, enabling good resolution
in the disk regions with strong density gradients and in the disk atmosphere.} This makes
it possible to couple our 2+1D model with compact chemical reaction networks to follow the long-term
chemodynamical evolution of young protostellar disks starting from the gravitational collapse 
of parental cloud cores. The details of this method will be presented in the upcoming paper.

{\it Acknowledgments.}
We are thankful to the anonymous referee for constructive comments that helped us to improve the manuscript
and the vertical reconstruction procedure.
This work was supported by the RFBR grant 17-02-00644.
The simulations were performed on the Vienna Scientific Cluster (VSC-2), on the Shared Hierarchical
Academic Research Computing Network (SHARCNET), on the Atlantic Computational Excellence Network 
(ACEnet). This publication is supported by the Austrian Science Fund (FWF).

\clearpage
\newpage
\begin{appendix}
\section{The thermal step and the reconstruction of the disk vertical structure}

The thermal step is placed in our algorithm between the source step (Equations~\ref{mom2_source} and
\ref{energ2_source}) and the transport step (Equations~\ref{cont2_adv}-\ref{energ2_adv}).
While the source and transport steps deal with two-dimensional quantities integrated 
over the disk vertical column 
(such as the gas surface density $\Sigma$, the vertically integrated pressure $\cal P$ and the internal
energy per surface area $e$) and are therefore defined on the $(r,\phi)$ polar mesh, 
the thermal step deals with three-dimensional quantities 
(such as the gas temperature $T$, the gas volume density $\rho$, the radiative energy $E$ and the
vertical column density from the disk midplane $\sigma$) defined on
the ($z,r,\phi$) cylindrical mesh.  
The initial values for all quantities (two- and three-dimensional) are provided by the initial setup
as described in Section~\ref{initial}.

Here, we describe the algorithm for solving Equations~(\ref{m1})
and (\ref{m2}), which take the disk radiative cooling/heating into
account. This algorithm also includes the reconstruction of the disk
vertical structure assuming the vertical hydrostatic equilibrium described by 
Equation~(\ref{static}). Our method is a
time-dependent modification to the steady-state models presented in
\citet{Akimkin:2012} and \citet{Vorobyov:2014}. The method for solving the 
source and transport steps are described in Section~\ref{solution} and in more 
detail in \citet{VB2010}.

Firstly, we note that the internal energy per surface area $e$ 
is updated in the source step due to 
the ${\cal P} (\nabla \cdot {\bl v})$ work and viscous heating. 
This may affect the gas temperature distribution in the disk and should be taken into account
before the thermal step is commenced. 
We assume that the generated (or consumed) heat in the source step is evenly redistributed 
over the disk vertical column so that the gas temperature in each element of the vertical column 
can be updated as
\begin{equation}
T^\ast = T^{n} {e^\ast \over e^{n}},
\label{tempr_update}
\end{equation} 
where index $n$ refers to the quantities at the beginning of the source step and
the asterix -- to the quantities after the source step. The updated 
gas temperature $T^\ast$ is further used in the thermal step.

In the second step, we compute the heating function $S$ (Equation~\ref{m1}) 
due to sources other than  the thermal radiation of the medium. We note that the dimension of $S$ by definition is $[\text{erg}\,
\text{s}^{-1}\, \text{g}^{-1}]$. In our model, these sources
include the UV radiation from the central star $S_{\text{star}}$ and
the interstellar UV radiation $S_{\text{bg}}$:
\begin{equation}
S=S_{\text{star}} + S_{\text{bg}},
\end{equation}
where $S_{\text{star}}$ is defined as 
\begin{equation}
S_{\text{star}} = 4\pi \kappa_{\text{P}}^{\text{star}} J_{\text{star}},
\end{equation}
where $\kappa_{\text{P}}^{\text{star}}$ is the mean Planck opacity
averaged over the stellar spectrum.

To compute $S_{\text{star}}$, we need
to know the distribution of the mean intensity of UV radiation
$J_{\text{star}}$. We calculate $J_{\text{star}}$
taking into account the absorption of the UV radiation by the disk as follows:
\begin{equation}
J_{\text{star}} = J^{0}_{\text{star}} \exp{\left( -\tau_{\text{star}}/\mu \right)},
\label{UV-RT}
\end{equation}
where $J^{0}_{\text{star}}$ is the UV intensity at the surface of the
disk, $\tau_{\text{star}}$ is optical depth calculated from the
surface of the disk,  and $\mu=\cos{\gamma_{\rm irr}}$ the cosine of the incidence angle of
stellar radiation arriving at the disk surface with respect to the
normal \citep[see][for details]{VB2010}. The optical depth as 
a function of the current column density $\sigma$ is defined as
\begin{equation}
\tau_{\text{star}}=\kappa_{\text{P}}^{\text{star}}
\left( \dfrac{\Sigma}{2} - \sigma \right),
\end{equation}
where $\sigma$ is measured from the disk mid-plane. The intensity
$J^{0}_{\text{star}}$ at the disk surface can be found as
\begin{equation}
J^{0}_{\text{star}} = \dfrac{1}{4\pi} \dfrac{L_\ast}{4\pi r^2}.
\end{equation}
Here, $L_\ast$ is the total 
stellar luminosity which includes contributions from photoshperic luminosity $L_{\rm ph}$
and accretion luminosity $L_{\rm accr}$ and $r$ the radial distance to the star.
The accretion luminosity is calculated as $L_{\rm acc}=G M_\ast \dot{M}/(2 R_\ast)$ using
information on the current stellar mass $M_\ast$, stellar radius $R_\ast$, and accretion
rate onto the star $\dot{M}$. The Photospheric luminosity and stellar radius are found from
the Lyon stellar evolution code coupled to the main hydrodynamics code as described 
in \citet{VB2015} and \citet{Baraffe2017}.

The background heating function $S_{\text{bg}}$ is calculated
by analogy to Equation~(\ref{UV-RT}), but with a
fixed value of $\mu=0.5$. As a boundary condition for the interstellar
radiation, we use the following relation:
\begin{equation}
J^{0}_{\text{bg}} = D\dfrac{caT_{\text{bg}}^4}{4\pi},
\end{equation}
where $c$ is the speed of light, $a$ the radiative constant,
$T_{\text{bg}}$ and $D$ the temperature and dilution of the
interstellar radiation. In our simulations, we adopt
$10\ 000$~K and $8\times 10^{-15}$, correspondingly.

In the third step, we calculate the change of the gas temperature $T$
and radiative energy $E$ in a given vertical column of the disk due to heating by 
the stellar UV and background radiation and cooling by the disk infrared emission. 
The thermal evolution of the vertical column is described by the system of radiative
transfer Equations~(\ref{m1}) and (\ref{m2}).

This system is a set of moment equations for diffuse IR
radiation derived under the Eddington approximation. We solve equations
\eqref{m1} and \eqref{m2} numerically to find $T$ and $E$ after one hydrodynamical time
step $\Delta t$ using the following implicit finite-difference scheme:
\begin{eqnarray}
&&c_{V} \dfrac{T-T^{*}}{\Delta t} = 
\kappa_P c (E-aT^4) + S \label{fd1} \\
&&\dfrac{E-E^{n}}{\Delta t} - \hat{\Lambda} E = -\rho^{n} \kappa_P c(E-aT^4),
\label{fd2}
\end{eqnarray}
where  $T^{*}$ is the gas temperature updated after the
source step (see Equation~\ref{tempr_update}), and
$E^{n}$ and $\rho^{n}$ the radiation energy and gas volume density taken 
from the previous time step.  In the discretized equations, we use the following convention:
index $i$ refers to the left-hand-side grid interface and index $i+1/2$ - to the grid center. 
In the above system, all quantities are defined at the grid center $i+1/2$ (the index omitted 
for simplicity) and $\hat{\Lambda} E$ denotes the finite-difference
approximation to the diffusion term:
\begin{eqnarray}
\hat{\Lambda} E = \dfrac{1}{\Delta z_{i+1/2}}
&&\left(\dfrac{c}{3\rho^{n}_{i+1}\kappa_{{\rm R},i+1}}\dfrac{E_{i+3/2}-E_{i+1/2}}{\Delta z_{i+1}}\right.\notag \\
&&\left.-\dfrac{c}{3\rho^{n}_{i}\kappa_{{\rm R},i}}\dfrac{E_{i+1/2}-E_{i-1/2}}{\Delta z_{i}} \right),
\end{eqnarray}
where
\begin{eqnarray*}
&&\Delta z_{i} = z_{i+1/2}-z_{i-1/2}\\
&&\Delta z_{i+1/2} = z_{i+1}-z_{i} \\
&&\rho^{n}_{i}\kappa_{{\rm R},i}=\dfrac{1}{2}\left(\rho^{n}_{i-1/2}\kappa_{{\rm R}.i-1/2}+\rho^{n}_{i+1/2}
\kappa_{{\rm R},i+1/2}\right).
\end{eqnarray*}
The above non-linear system is solved with the iterative Newton method.
Namely, we approximate $T^4$ as $(T^{k})^4
\left(\dfrac{4T}{T^{k}}-3\right)$ using the first two terms of the Taylor expansion series,
where superscript $k$ refers to the
previous iterative step and produce a system of linear equations 
for $E_{1/2},...,E_{M-1/2}$ with a tridiagonal matrix, where $M$ is the number of grid cells in the
vertical direction. This tridiagonal
system is solved with the forward and back substitution method (the Thomas
algorithm). The obtained values of energies are back substituted in Equation~(\ref{fd1}) to derive
the temperatures and both quantities are then used for the next Newton iteration as 
$T^{k+1}$ and $E^{k+1}$ until convergence is achieved. The solution is then set to $T=T^{k+1}$ 
and $E =E^{k+1}$.

The boundary conditions for Equations~(\ref{fd1}) and (\ref{fd2}) are the zero energy gradient
near the mid-plane and following relation between the diffusion flux
and radiation energy at the disk surface:
\begin{equation}
-\dfrac{c}{3\rho^{n}\kappa_\text{R}} \left.\dfrac{\partial E}{\partial z} \right|_{z_{M}}
=\dfrac{c}{2} \left(E - aT_\text{cmb}^4\right),
\label{ir_out}
\end{equation}
where $T_\text{cmb}$=2.73~K. The latter condition is derived under the
assumption that the incoming and outgoing radiation is isotropic over the
upper and lower hemispheres. We note that the presented method is
conceptually similar to that which is developed for the calculation of
the pre-stellar core thermal evolution in \cite{Pavlyuchenkov:2015}.

The next stage of the algorithm is to recover the hydrostatic
equilibrium along the vertical direction using the updated temperatures. 
Equation~(\ref{static}) describing
the vertical hydrostatic equilibrium can be written as follows:
\begin{eqnarray}
&&\dfrac{dz}{d\sigma} = \dfrac{1}{\rho} \label{sigma}\\
&&{R \over \mu }\dfrac{d(\rho T)}{d\sigma} =  -\dfrac{GM_\ast}{r^3}z - 4\pi G \sigma,
\label{static_app}
\end{eqnarray}
where $r$ is the
radial distance to a given vertical column of gas, $G$ the gravitational constant,
$\sigma$ the gas column density measured from the disk mid-plane. We note
that $T$ is known from the previous step. Equation~\eqref{static_app}
accounts for the gravity of the central star and disk self-gravity in the plane-parallel limit. 
The boundary conditions for this system have the following form:
\begin{eqnarray}
&&z(0)=0 \\
&&\rho(\sigma_{M}) = \rho_{ext},
\end{eqnarray}
where $\rho_{ext}$ is the gas volume density at the disk surface.
In our models, $\rho_{ext}$  corresponds to a hydrogen
number density of $10^3$~cm$^{-3}$.

The solution of the system \eqref{sigma} --- \eqref{static_app}
with the corresponding boundary conditions can be found using
an implicit scheme similar to that used when
computing the internal structure of the stars. We
linearize the right-hand side of $\eqref{sigma}$ as follows:
\begin{equation}
\dfrac{1}{\rho} \approx \dfrac{1}{\rho^{k}} - \dfrac{1}{(\rho^{k})^2}
(\rho-\rho^{k}),
\end{equation}
which transforms the initial system into a linear system of ordinary
differential equations. The implicit finite-difference approximation
of the latter one generates a system of algebraic linear equations
with a tridiagonal matrix whose solution can be easily found using the
forward and backward substitution method. The resulting solution is used to
form a new approximation, and iterations over $k$ are carried out until
convergence is achieved (usually, after a few iterations).

\begin{figure}
\centering
\includegraphics[width=0.45\textwidth]{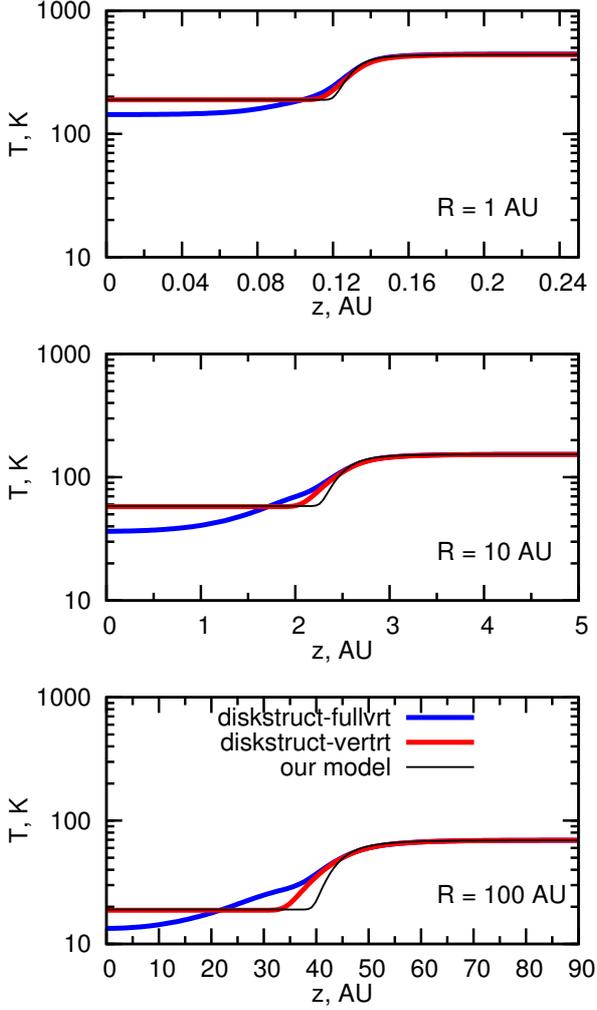}
\caption{Vertical temperature distributions of a protoplanetary
disk calculated with our method and with
the ``diskstruct'' package written by C. Dullemond. More specifically, 
the results from the ``diskstruct-fullrt'' method with the full treatment of
the frequency and angular dependent radiative transfer are shown with
the thick blue lines. The thick red curves represent the results from the approximate
``diskstruct-vertrt'' method, which adopts grey opacities for
the dust thermal radiation. The temperature distributions calculated with
our method are shown with the thin black lines. The presented vertical
distributions are calculated for three radial positions in the disk, namely for
$R=1$~AU (top), $R=10$~AU (middle),  and $R=100$~AU (bottom) for 
an optically thick protoplanetary disk illuminated
by a star with $T_\text{eff}=6000$~K.}
\label{diskstruct}
\end{figure}

At the last step of the algorithm, we calculate the updated thermal
energy $e$ per unit area by summing up the thermal energies over the
vertical grid:
\begin{equation}
e = 2 c_{V}\sum\limits_{k=0}^{M} T_{i+1/2} \left(\sigma_{i+1}-\sigma_{i}\right), 
\label{energy2D}
\end{equation}
where a factor of 2 accounts for the fact that we adopted an equatorial symmetry
with resect to the disk midplane.
This value will be further used in the transport step.

Finally, we note that the transport step updates the values of $\Sigma$ and 
$e$ (see Equations~\ref{cont2_adv} and \ref{energ2_adv}), which in turn affects 
the volumetric distributions of the gas temperature $T$ 
and volume density $\rho$. To take these changes into account,  we assume that the mass and 
thermal energy that are carried with the flow are evenly
redistributed over the vertical cells so that the updated distributions of $T$
and $\rho$ can be calculated as:
\begin{eqnarray}
%&&\sigma^{*}_{i} = \sigma^{n}_{i}\dfrac{\Sigma^{*}}{\Sigma^{n}} \\
&&\rho^{n+1} = \rho\dfrac{\Sigma^{n+1}}{\Sigma}, \\
&& T^{n+1} = T \dfrac{\Sigma}{\Sigma^{n+1}} \dfrac{e^{n+1}}{e},
\end{eqnarray}
where $T$, $\rho$,  and $\Sigma$ are the values of the gas temperature, 
its volume and surface densities before the transport step. This completes
one cycle of integration and the updated values $T^{n+1}$, $\rho^{n+1}$,  
and $\Sigma^{n+1}$ are used at the next time step.

Since the method is fully implicit, it is in general very stable.
However, in very few cases the Newton procedure may not converge for
some values of the time step. In this case, we divide the hydrodynamic 
time step into several sub-cycles,
which usually solves the problem. In the limit of large time steps, the method
yields a steady-state solution similar to that obtained
by the steady-state model described in \cite{Vorobyov:2014}.

The presented algorithm was carefully tested and compared with other
methods. In particular, we benchmark our steady-state solutions (in the limit of large time steps) 
with the results of 1+1D code 
``diskstruct''\footnote{http://www.ita.uni-heidelberg.de/$\sim$dullemond/\\
software/diskstruct/index.shtml}
developed by C. Dullemond and discussed in \cite{Dullemond:2002}. The
results of this comparison are shown in Figure~\ref{diskstruct}. The
vertical temperature distributions calculated with our method turn out 
to be very similar to the results obtained with the approximate method
``vertrt'', which also adopted grey opacities for the dust thermal radiation 
included in the ``diskstruct'' package. The deviation of our solution from
the results produced by the more accurate method ``fullrt'',  which takes into
account the frequency and angular dependence of radiative transfer, is
notable near the equatorial plane. However, we consider these
differences to be acceptable for our simulations.

\begin{figure}
\centering
\includegraphics[width=0.45\textwidth]{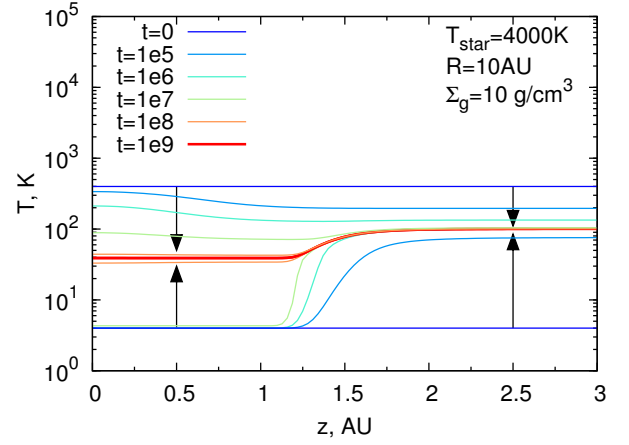}
\caption{Evolution of the gas temperature distribution at $r=10$~AU in
a toy model of the protoplanetary disk. The density structure is fixed.
The initial temperature distribution is uniform with $T=4$~K (first
case) and $T=400$~K (second case). The initial distributions are shown
with the shades of blue, the steady-state solution is shown with the red color.
The arrows illustrate the evolution of the gas temperature profiles with time. The corresponding time in seconds is shown in the color legend. }
\label{T_evol}
\end{figure}
To illustrate the time-dependent aspect of the
radiative transfer model used in the thermal step, we show in Figure~\ref{T_evol} 
the evolution of the temperature
distribution at $r=10$~AU with the density distribution manually fixed. 
For the parameters of the considered vertical column of gas,
the relaxation to a steady-state solution is achieved within about
a few years (about an order of magnitude faster than the dynamical time at $r=10$~AU)
starting from either a warmed-up disk (with an
initial uniform temperature of $T=400$~K) or from a cooled-down disk 
(with a uniform temperature of $T=4$~K). This relaxation time turns out to be
in agreement with the characteristic thermal time estimated
in \citet{Vorobyov:2014}. As expected, the relaxation to the steady-state
solution is faster in the upper layers of the disk where the gas volume density
is lower.

\begin{figure}
\centering
\includegraphics[width=0.45\textwidth]{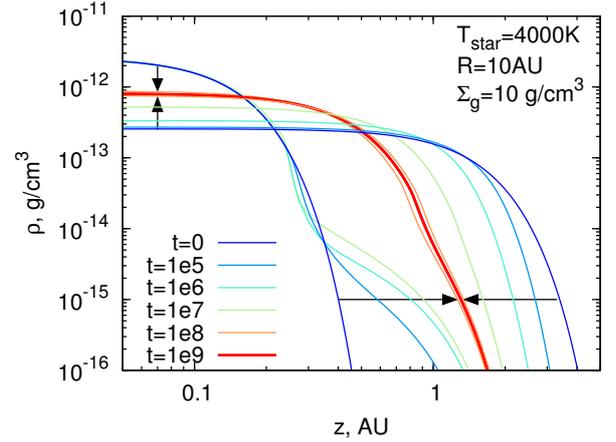}
\caption{Evolution of the gas volume density distribution at $r=10$~AU in
a model of protoplanetary disk. The initial temperature distribution
is uniform with $T=4$~K (first case) and $T=400$~K (second case). The
initial density distributions are shown with the shades of blue, the
steady-state solution is shown with the red color. 
The arrows illustrate the evolution of the gas density profiles with time.
The corresponding time in seconds is shown in the color legend.  }
\label{Rho_evol}
\end{figure}
While producing the temperature distributions in Figure~\ref{T_evol}, 
the gas volume density distribution was manually fixed. When the density is allowed to 
evolve together with the temperature, then the density distribution develops as shown in
Figure~\ref{Rho_evol}. We see that at low and high initial temperatures
the disk vertical height is low and high, respectively. In the steady-state
profile near $z=1.2$~AU one can see a slight change in the slope
which corresponds to the change of the temperature profile.

\end{appendix}
\end{document}